\documentclass[12pt]{article}
\usepackage{times}
\usepackage{geometry}
\usepackage[utf8]{inputenc}
\usepackage{enumitem,amssymb}
\usepackage{ragged2e}
\usepackage{fancyhdr}
\usepackage[sort,numbers]{natbib}
\usepackage{graphics,graphicx}
\usepackage{color}
\usepackage{xcolor}
\usepackage{lineno}
\usepackage{geometry}
\usepackage{wasysym}

\usepackage{adjustbox}
\usepackage{array}

\usepackage[compact]{titlesec}

\newcolumntype{R}[2]{%
    >{\adjustbox{angle=#1,lap=\width-(#2)}\bgroup}%
    l%
    <{\egroup}%
}
\newcommand*\rot{\multicolumn{1}{R{90}{0em}}}

\newlist{thematic}{itemize}{8}
\setlist[thematic]{label=$\square$}
\usepackage{pifont}
\newcommand{\cmark}{\ding{51}}%
\newcommand{\done}{\rlap{$\square$}{\raisebox{2pt}{\large\hspace{1pt}\cmark}}%
\hspace{-2.5pt}}

\begin{document}
\raggedright
\huge
Astro2020 Science White Paper \linebreak

Opportunities for Multimessenger Astronomy in the 2020s \linebreak
\normalsize

\noindent \textbf{Thematic Areas:} \hspace*{60pt} $\square$ Planetary Systems \hspace*{10pt} $\square$ Star and Planet Formation \hspace*{20pt}\linebreak
$\square$ Formation and Evolution of Compact Objects \hspace*{31pt} $\square$ Cosmology and Fundamental Physics \linebreak
  $\square$  Stars and Stellar Evolution \hspace*{1pt} $\square$ Resolved Stellar Populations and their Environments \hspace*{40pt} \linebreak
  $\square$    Galaxy Evolution   \hspace*{45pt} $\done$             Multi-Messenger Astronomy and Astrophysics \hspace*{65pt} \linebreak
  
\textbf{Principal Author:}

Name:	Eric Burns
 \linebreak						
Institution:  NASA Goddard
 \linebreak
Email: eric.burns@nasa.gov
 \linebreak
Phone:  +1-301-286-4664
 \linebreak

\textbf{Co-authors:} 

\noindent A. Tohuvavohu (Penn State University), J. M. Bellovary (Queensborough Community College, American Museum of Natural History), E. Blaufuss (University of Maryland), T. J. Brandt (NASA Goddard), S. Buson (University of W{\"u}rzburg, University of Maryland Baltimore County), R. Caputo (NASA Goddard), S. B. Cenko (NASA Goddard), N. Christensen (Observatoire de la C{\^o}te d’Azur), J. W. Conklin (University of Florida), F. D'Ammando (INAF-IRA Bologna), K.E.S. Ford (CUNY Borough of Manhattan Community College, American Museum of Natural History), A. Franckowiak (DESY), C. Fryer (LANL), C. M. Hui (NASA Marshall), K. Holley-Bockelmann (Vanderbilt), T. Jaffe (University of Maryland, NASA Goddard), T. Kupfer (UC Santa Barbara), M. Karovska (CFA), B. D. Metzger (Columbia University), J. Racusin (NASA Goddard), B. Rani (NASA Goddard), M. Santander (University of Alabama), J. Tomsick (UC Berkeley), C. Wilson-Hodge (NASA Marshall)

\justify

\noindent \textbf{Abstract:}

\noindent Electromagnetic observations of the sky have been the basis for our study of the Universe for millennia, cosmic ray studies are now entering their second century, the first neutrinos from an astrophysical source were identified three decades ago, and gravitational waves were directly detected only four years ago. Detections of these messengers are now common. Astrophysics will undergo a revolution in the 2020s as multimessenger detections become routine. The 8th Astro2020 Thematic Area is Multimessenger Astronomy and Astrophysics, which includes the identification of the sources of gravitational waves, astrophysical and cosmogenic neutrinos, cosmic rays, and gamma-rays, and the coordinated multimessenger and multiwavelength follow-ups. Identifying and characterizing multimessenger sources enables science throughout and beyond astrophysics. Success in the multimessenger era requires: (i) sensitive coverage of the non-electromagnetic messengers, (ii) full coverage of the electromagnetic spectrum, with either fast-response observations or broad and deep high-cadence surveys, and (iii) improved collaboration, communication, and notification platforms.

\pagebreak

\section{Introduction}
Astrophysical observatories now detect several messengers from the Universe:
\begin{itemize}[noitemsep,topsep=0pt]
    \item \textbf{Photons} are electromagnetic (EM) radiation. We now observe photons from radio wavelengths up to very high energy gamma-rays, covering more than twenty decades in energy.
    \item \textbf{Gravitational waves (GWs)} are spacetime ripples emitted from systems with an accelerating quadrupole moment. Detectable GWs are expected from binary systems of white dwarfs (WDs), neutron stars (NSs), and black holes (BHs) or possibly from non-axisymmetric dense objects such as a non-spherical NSs, cosmic strings, or core-collapse supernovae.
    \item \textbf{Neutrinos} are the lightest massive particles and are produced in weak interactions. In astrophysics we expect $\sim$MeV neutrinos from core-collapse events and high-energy ($\sim$TeV-PeV) neutrinos from efficient particle reservoirs/accelerators such as supernova remnants or relativistic jets from active galactic nuclei.
    \item \textbf{Cosmic Rays} are high-energy charged particles such as protons or atomic nuclei. Heavier nuclei are formed and released in explosive nucleosynthesis, and achieve high kinetic energies from natural particle accelerators.
\end{itemize}

\noindent In astronomy, giant leaps forward have followed the opening of new observations of the EM spectrum. As multiwavelength studies brought new understanding, multimessenger observations will revolutionize our field as each of the different messengers carries distinct information that can be combined for a fuller understanding. Critical to future multimessenger science are joint observations, multiwavelength EM coverage, and improved communication. We demonstrate this using prior multimessenger examples and give broad recommendations to maximize science in the new multimessenger era.

\section{The Past and Present}
Among the Astro2020 topics is the Coordinated Multimessenger and Multiwavelength follow-ups of multimessenger sources. We use the three historic convincing multimessenger transients in astronomy to demonstrate the necessary capabilities to achieve this goal. The science from these discoveries is counted in thousands of papers and summarized by Astro2020 topic in our Table. However, understanding multimessenger sources is not the only science possible through multimessenger studies. We also discuss a representative sample of such work. 

\subsection{The Core-Collapse Supernova SN 1987A}
The transient multimessenger era began 30 years ago with the detection of the nearest supernova in centuries: SN 1987A, a core-collapse supernova (CCSN) explosion in the Large Magellanic Cloud \citep{SN1987A_optical}. The LMC is an oft-studied object; this led to the early optical detection by chance, and the identification of the progenitor as the blue supergiant Sanduleak -69 202 \citep{SN1987A_sanduleak}. Independently, a burst of $\sim$MeV neutrinos was detected a few hours before optical identification \citep{SN1987A_neutrinos,SN1987A_neutrinos_2,SN1987A_neutrinos_3}. The unambiguous association of these two signals arose because of the all-sky monitoring for MeV neutrinos and the early optical observation of the explosion.  Multimessenger studies of this event confirmed that some supernovae are produced in the formation of NSs and BHs, gave new insight into the supernova engine, unexpectedly showed blue supergiants end in core-collapse, set upper bounds on the neutrino mass, charge, and number of flavors, and enabled some unique tests of gravity \citep[e.g.][]{SN1987A_gen,SN1987A_WEP,SN1987A_axions,SN1987A_dimensions}.



SN 1987A has been observed since its explosion, and the breadth of observations have provided a broad understanding of this event. However, due to a lack of observations of the shock break-out and a lack of sensitive coverage of MeV gamma-rays and GWs possible science was lost, some of which remains unknown. To capture the early emission for future events, the SuperNova Early Warning System (SNEWS) correlates MeV neutrino bursts from multiple observatories and rapidly distributes them to the astronomical community \citep{SNEWS}. With current instruments it is expected to trigger on a real event every few decades. It is critical to have serendipitous joint observations with $\sim$kHz GW observatories and fast-response capability across the electromagnetic spectrum.

\subsection{The Binary NS Merger GW170817, GRB 170817A, AT2017gfo}
The short gamma-ray burst GRB 170817A was detected by the \textit{Fermi} Gamma-ray Burst Monitor (GBM) on August 17th, 2017 \citep{GW170817_GBM}. Independently, GW170817 was identified in online searches of LIGO/Virgo strain data \citep{GW170817_LVC}. Both signals originated from the merging of two neutron stars in NGC 4993 \citep{GW170817-Standard-Siren}. The GW detection kicked off a worldwide follow-up campaign involving the suite of NASA astrophysics missions and dozens of ground-based observatories on every continent covering all messengers and the full range of the electromagnetic spectrum \citep{GW170817_MMAD}. The discovery of the early ultraviolet emission by the Neil Gehrels \textit{Swift} Observatory and temporal evolution of the afterglow (as informed from \textit{Swift}, \textit{NuSTAR}, \textit{Chandra}, and \textit{XMM} observations, with radio partners), provided key insight into the event. The multimessenger detection enabled the classification of the event as a binary NS (BNS) merger \citep{GW170817_MMAD}, confirmed that BNS mergers are progenitors of both short GRBs and kilonovae, allowed the first direct identification of an astrophysical source of the heaviest (r-process) elements, set new constraints on the equation of state of supranuclear matter, and measured the speed of gravity.

The event was a remarkable success. Decades of investment by the NSF into GW interferometers, the development of real-time search and reporting pipelines, and follow-up with dozens of EM and neutrino facilities resulted in a discovery that affected several fields beyond the astrophysics of the source itself. Future events will require faster identification of the source position and multiwavelength coverage to resolve the unsettled questions. This has led the GBM and LIGO/Virgo teams to build an automated multimessenger association and reporting pipeline to facilitate success in follow-up observations.


\subsection{The Blazar TXS 0506+056}
IceCube has developed real-time alerts for candidate astrophysical neutrino events. Neutrinos in the range observed by IceCube are expected to track X-ray emission. \textit{Swift} performs rapid follow-up of the IceCube alerts in the soft X-ray regime, which led to the identification of 9 X-ray sources in the region of IceCube-170922A \citep{TXS_IceCube}, one of which was the blazar TXS 0506+056 \citep{TXS_Swift}. About a week later, the \textit{Fermi}-LAT Collaboration announced that TXS 0506+056 was in middle of the brightest $\sim$GeV flare in a decade of observations, beginning several months before the neutrino arrival \citep{TXS_Fermi}. This triggered a large follow-up effort, resulting in the first detection of this source at very-high energies by MAGIC \citep{TXS_MAGIC}. Though the association is not entirely unambiguous, these observations provide compelling evidence for the identification of the first individual source of high-energy neutrinos, the first extragalatic neutrino source, a potential source of cosmic rays, and new insights into the composition of jets from active galactic nuclei (AGN) \citep{TXS_overview}.

This event was again a success for coordinated follow-up observations. The real-time search and reporting of IceCube neutrinos, the all-sky monitoring by \textit{Fermi}-LAT, and the fast response of \textit{Swift}, MAGIC, and other follow-up observatories resulted in a quick identification of an interesting transient and broadband characterization. In response to this event, improvements have been made to follow-up neutrino searches, \textit{Fermi}-LAT transient reporting, and follow-up procedures by \textit{Swift} and other observatories. 


\subsection{Broader Multimessenger Studies}
Multimessenger Astronomy is much broader than transients, and even the four messengers that fall under the scope of the Astro2020 Decadal. We list a few examples to show the breadth of science multimessenger studies enable. Dust grains from ocean sediments or meteorites have been used to determine isotopic ratios to understand supernova \citep{dust_1,dust_2} and to argue that NS mergers are important production sites for heavy elements \citep{dust_3}. 
X-ray and radio observations of supernova remnants (SNRs) identified them as likely sources of cosmic rays. Combining this with gamma-ray observations and particle physics theory gave the first evidence of proton acceleration in SNRs \citep{LAT_cosmic_rays_SNR}. Because propagation affects some messengers differently, combining observations of photons and cosmic rays enables new ways to study the Intergalactic and Galactic Magnetic Field (GMF), which will enable a greater understanding of cosmic ray propagation and better enable their use in other multimessenger studies \citep{GMF}.

\section{The Future}
We here make general recommendations to ensure success in the multimessenger era. Example Astro2020 white papers with more detailed descriptions of exciting multimessenger science include: NS mergers \citep{WP_NS_mergers}, CCSN \citep{WP_CCSN,LANL_CCSN}, Blazars/Active Galactic Nuclei (AGN) \citep{WP_AGN,WP_AGN_2}, supermassive black hole binaries (SMBHB) \citep{WP_SMBHB_1,WP_SMBHB_2,NatarajanWP}, intermediate mass BHs (IMBHs) \citep{IMBHs}, galactic binaries \citep{WP_galactic_binaries,galactic_binaries_2}, stellar-mass binary black hole (sBBH) mergers \citep{WP_SBBH_1,WP_SBBH_2,CutlerWP}, studies of the GMF \citep{GMF}, and searches for the origin of new messengers \citep{WP_UHECRs_fundamental,WP_UHECRs,CR_venters,WP_UHE_general,BakerWP,EracleousWP}. A broad, but incomplete, summary of multimessenger science and the necessary capabilities required to uncover them is given in the Table. Below we summarize the current and forseeable state of observational coverage of these messengers.





\subsection{Detecting GWs, Neutrinos, Cosmic Rays, and Gamma-rays}
The ground-based GW network covers the $\sim$10-1000 Hz range. The current network is approach its design sensitivity and an upgrade to the LIGO interferometers has been funded (expected $\sim$2025). The European-led LISA mission is a space-based interferometer sensitive to the $\sim$0.1-100 mHz frequency range (launch expected $\sim$2014). Pulsar Timing Arrays look for lower frequency GWs in the $\sim$1-1,000 nHz range, corresponding to orbital periods of weeks to decades.

Several MeV neutrino detectors exist, with 7 integrated into SNEWS. In the $\sim$TeV-PeV regime, the IceCube Upgrade is underway and Km3Net is under construction. New cosmic ray detectors have been launched in recent years, including AMS-02, CALET, and SuperTIGER. The largest ground-based observations for ultra-high energy cosmic rays are the Pierre Auger Observatory and the Telescope Array with are being upgraded. The active gamma-ray surveys are \textit{Fermi} and HAWC, covering 0.04-300 GeV and 0.1-100 TeV, respectively. These work in concert with the sensitive narrow-field Imaging Atmospheric Cherenkov Telescopes (IACTs) such as H.E.S.S., MAGIC, and VERITAS. The international community is gearing up for the full construction of the Cherenkov Telescope Array that promises vast improvements over the current generation IACTs.

There are proposed upgrades to several of these existing missions, and proposed large-scale missions that promise vast improvements to our ability to directly detect each messenger and identify their sources. The sources of GWs, neutrinos, cosmic rays, and gamma-rays cannot be identified without the ability to observe each messenger directly. Determining the astrophysical relationship of signals from multimessenger sources requires spatial and temporal information, and sometimes contemporaneous observations. \textit{We thus broadly recommend continued and capable coverage to missions that detect these messengers.}

\subsection{EM Observations}
Multimessenger studies of NS mergers, CCSN, and AGN are all best informed with observations across the EM spectrum. Many sources are explosive transients or compact objects, which often have thermal and non-thermal emission. The three historical examples also show the importance of time in multimessenger astronomy, with science sometimes requiring serendipitous joint observations via all-sky monitoring, high-cadence surveys, and fast-response follow-up over the entire EM spectrum. This capability has led to surprise discoveries when met, and a loss of knowledge when not. 


New radio telescopes, surveys, and interferometry have brought new capabilities and coverage, with larger-scale missions being proposed. Infrared will soon be covered by three flagship missions, and near-infrared from ground-based observations. Optical coverage now has several transient surveys, and the forthcoming Extremely Large Telescopes and LSST promise to reliably cover these bands. CTA promises improved sensitivity for pointed TeV observations. \textit{We recommend appropriate resources be made available for multimessenger studies with these instruments. Gamma-rays in the $\sim$keV-GeV energy range, X-rays, and ultraviolet light can only be observed from space and can be crucial for multimessenger studies; we recommend continuous and sensitive survey or fast-response coverage in these energies.}

\subsection{Coordinated Multimessenger and Multiwavelength Follow-ups}
Response time can be absolutely critical for a full understanding of astrophysical events. For this purpose, many time-domain discovery missions have created real-time alert pipelines. In the case of GW observations of inspiraling compact objects and MeV neutrino observations of CCSN, the detections of these signals can be reported before the first light is detectable, though the localizations generally require wide-field partner observatories. Continued improvements to real-time communication methods and coordination is paramount. This includes reducing the response time of space-based observatories, generally limited by uplink delay. Enabling inter-mission communication and coordination requires devoted resources. As the field has grown, funding mechanisms have not caught up. This is, in part, caused by the NSF and NASA separation as they both have critical assets for multimessenger astronomy. \textit{We recommend improved collaboration, communication, and notification platforms. Further, we recommend that all relevant agencies coordinate their plans and goals for multimessenger astronomy.}

\begin{table}
\centering
\footnotesize
\begin{tabular}{|l|c|c|c|c|c|c|c|c|c|c|}
\hline
Messengers; \textbf{Astro2020 Thematic Areas} and Sub-Topics                                                              & \rot{NS Mergers} & \rot{CCSN} & \rot{AGN/Blazars} & \rot{SMBHBs} & \rot{IMBHs} & \rot{Gal. Binaries} & \rot{Pulsars} & \rot{sBBH} & \rot{SNRs} & \rot{GMF} \\ \hline
GWs - kHz                                                                  & x          & ?    &             &       &      &               & x       & x    &      &     \\
GWs - mHz                                                                  &            &      & ?           & x     & x    & x             &         & x    &      &     \\
GWs - nHz                                                                  &            &      & ?           & x     &      &               &         &      &      &     \\ \hline
Neutrinos - TeV-EeV                                                        & x          & x    & x           &       &      &               &         & ?    & x    &     \\
Neutrinos - MeV-GeV                                                        & x          & x    &             &       &      &               &         &      & x    &     \\ \hline
Cosmic Rays - Ultra High Energy                                            & x          & x    & x           &       &      &               &         &      &      & x   \\
Cosmic Rays - High Energy                                                  & x          & x    & x           & x     &      &               &         &      & x    & x   \\ \hline
Gamma-rays - keV-TeV                                                       & x          & x    & x           & ?     & ?    & x             & x       & ?    & x    & x   \\
X-rays                                                                     & x          & x    & x           & x     & x    & x             & x       & x    & x    &     \\
UV                                                                         & x          & x    & x           & x     & x    & x             & x       & x    & x    &     \\
Optical                                                                    & x          & x    & x           &       & x    & x             & x       & x    & x    & x   \\
IR                                                                         & x          & x    & x           &       & x    & x             & x       & x    & x    & x   \\
Radio                                                                      & x          & x    & x           & x     & x    & x             & x       & x    & x    & x   \\ \hline\hline
\textbf{Thematic Area 2: Star and Planet Formation}                        &            &      &             &       &      &               &         & x    & x    & x   \\ 
Formation of Stars and Clusters                                            &            &      &             &       &      &               &         & x    & x    &     \\
Molecular Clouds and the Cold Interstellar Medium; Dust                    &            &      &             &       &      &               &         &      & x    & x   \\ \hline
\textbf{Thematic Area 3: Stars and Stellar Evolution}                      & x          & x    &             &       &      & x             & x       & x    & x    &     \\ 
Stellar Astrophysics                                                       &            & x    &             &       &      & x             & x       & x    & x    &     \\
Structure and Evolution of Single and Multiple Stars                       & x          & x    &             &       &      & x             & x       & x    & x    &     \\ \hline
\textbf{Thematic Area 4: Formation and Evolution of Compact Objects}       & x          & x    &             &       & x    & x             & x       & x    & x    &     \\ 
Stellar-mass Black Holes                                                   & x          & x    &             &       &      & x             &         & x    & x    &     \\
Neutron Stars                                                              & x          & x    &             &       &      & x             & x       &      & x    &     \\
White Dwarfs                                                               &            &      &             &       &      & x             &         &      & x    &     \\
Supernovae                                                                 &            & x    &             &       &      & x             &         &      & x    &     \\
Mergers of Compact Objects                                                 & x          &      &             &       & x    &               &         & x    & x    &     \\
Gamma-ray Bursts                                                           & x          & x    &             &       &      &               &         &      &      &     \\
Accretion                                                                  & x          & x    &             &       &      & x             & x       & x    &      &     \\
Production of Heavy Elements                                               & x          & x    &             &       &      &               &         &      & x    &     \\
Extreme Physics on Stellar Scales                                          & x          & x    &             &       &      & x             & x       & x    & x    &     \\ \hline
\textbf{Thematic Area 5: Resolved Stellar Populations/Environments}        &            &      &             &       & x    & x             & x       & x    & x    & x    \\ 
Structure and Properties of the Milky Way and Nearby Galaxies              &            &      &             &       & x    & x             &         &      & x    & x   \\
Stellar Populations and Evolution                                          &            &      &             &       &      & x             & x       & x    & x    &     \\
Interstellar Medium and Star Clusters                                      &            &      &             &       &      &               &         & x    & x    & x   \\ \hline
\textbf{Thematic Area 6: Galaxy Evolution}                                 &            &      & x           & x     & x    &               &         & x    & x    & x    \\ 
(Forma/Evolu)tion/Dynamics/Properties of SMBHs/Galaxies/Clusters  &            &      & x           & x     & x    &               &         & x    &      & x   \\
Active Galactic Nuclei and QSOs                                            &            &      & x           & x     &      &               &         & x    &      &     \\
Mergers                                                                    &            &      & x           & x     & x    &               &         &      &      &     \\
Star Formation Rates                                                       &            &      & x           &       &      &               &         & x    & x    &     \\
Gas Accretion; Circumgalactic and Intergalactic Media                      &            &      & x           & x     &      &               &         &      & x    & x   \\ \hline
\textbf{Thematic Area 7: Cosmology and Fundamental Physics}                & x          & x    & x           & x     &      &               & x        & x    & x    & x    \\ 
Early Universe                                                             &            &      & x           &       &      &               &         &      &      &     \\
Cosmic Microwave Background                                                &            &      & x           &       &      &               &         &      &      & x   \\
Determination of Cosmological Parameters                                   & x          &      & x           & x     &      &               &         & x    &      &     \\
Dark Matter and Dark Energy                                                & x          &      &             & x     &      &               &         &      &      &     \\
Astroparticle Physics                                                      & x          & x    & x           & x     &      &               & x       &      & x    & x   \\
Tests of Gravity                                                           & x          &      & x           & x     &      &               & x       & x    &      &     \\
Astronomically Determined Physical Constants                               & x          &      & x           & x     &      &               &         &      &      &     \\ \hline
\textbf{Thematic Area 8: Multi-Messenger Astronomy and Astrophysics}       & x          & x    & x           & x     & x    & x             & x       & x    & x    & x   \\ 
Identify Sources of GWs, Neutrinos, Cosmic Rays, and Gamma-rays        & x          & x    & x           & x     & x    & x             & x       & x    & x    & x   \\
Coordinated Multimessenger and Multiwavelength Follow-ups                  & x          & x    & x           & x     & x    & x             & x       & x    & x    & x   \\ \hline
\end{tabular}\label{tab:summary}
\end{table}

\section{Summary}
With multimessenger astrophysics, we will unlock otherwise inaccessible secrets of the Universe. The three prior multimessenger events have brought unprecedented understanding throughout astrophysics. Maximizing multimessenger science requires a broad and balanced portfolio and facilitation of inter-mission coordination and development.


\pagebreak

\bibliographystyle{abbrv}
\bibliography{references}

\end{document}